# Nonlinear switching of low-index defect modes in photonic lattices

Fangwei Ye[1], Yaroslav V. Kartashov[1], Victor A. Vysloukh[2], and Lluis Torner[1]

[1]*ICFO-Institut de Ciencies Fotoniques, and Universitat Politecnica de Catalunya, Mediterranean Technology Park, 08860 Castelldefels (Barcelona), Spain*

[2]*Departamento de Fisica y Matematicas, Universidad de las Americas – Puebla, Santa Catarina Martir, 72820, Puebla, Mexico*

We address nonlinear signal switching between two low-index defect channels induced in periodic optical lattices. In contrast to conventional directional couplers, where the guiding mechanism is total internal reflection or refraction, in such Bragg-type coupler, the guidance is of a photonic-bandgap origin. The coupling length in the low-index coupler is controlled by the lattice parameters and by the channel spacing. In the nonlinear regime the Bragg- type coupler behaves as an all-optical switch, exhibiting a remarkable difference of switching power for focusing versus defocusing nonlinearity.

Photonic bandgap structures (PBG), including photonic crystals [1] and photonic crystal fibers (PCF) [2-4], have attracted an enormous interest during the last decade. Attention has typically focused on high-contrast refractive-index structures. More recently, low-contrast optically induced nonlinear PCF-like structures were introduced [5]. Importantly, in such low-contrast lattices light gets localized due to distributed Bragg-type backscattering, which involves at least a few lattice periods. This is in contrast to high-contrast microstructured geometries where anti-resonant reflecting mechanism frequently dominates and the transmission spectrum is determined by the index contrast and the thickness of the first high-index layer rather than by the lattice period [6]. The linear and nonlinear guiding properties of defects imprinted in such shallow lattices have been extensively discussed [7-9]. In particular, the possibility of linear directional bandgap coupling, which leads to periodic energy switching between defects akin to switching between the cores in conventional directional couplers [10-15], was put forward for one-dimensional lattices with low-index defects [9]. Nonlinear effects have been studied mostly in dual-core PCFs with high-index defects [16,17]. Photonic band-gap couplers are more flexible than the conventional couplers, because of the possibility of engineering their dispersion characteristics by changing the lattice parameters and by using lattices with amplitude/frequency modulation. An enhanced flexibility is also afforded by the opportunity to operate with modes belonging to different gaps and by the potential switching of higher-order modes. Note also that photonic bandgap nanoscale structures are suitable for guiding of surface plasmon polaritons [18].

In this paper we study the impact of the nonlinearity on the coupling and switching characteristic between two low-index defect channels imprinted in a periodic harmonic lattice. We find that periodic power coupling between the low-index defects is possible only for non-overlapping domains of the lattice modulation depths. A remarkable difference of switching powers in focusing and defocusing media is a characteristic feature of Bragg-type coupler that we also analyze. The possibility of switching with higher-order (dipole-like) modes is shown.

We start from the governing equation for the complex amplitude of the field $q$ of a light beam propagating in a medium with a shallow transverse refractive index modulation that can be derived from Maxwell equations for a dielectric medium in which one assumes a cubic nonlinearity. The derivation procedure is similar to that used in ref [19], where one uses the slowly varying amplitude approximation to arrive at the following nonlinear Schrödinger equation:

$$i\frac{\partial q}{\partial \xi} = -\frac{1}{2}\frac{\partial^2 q}{\partial \eta^2} + \sigma |q|^2 q - pR(\eta)q. \qquad (1)$$

Here the transverse $\eta$ and longitudinal $\xi$ coordinates are scaled to the characteristic beam width and diffraction length, respectively; the parameter $p$ characterizes the lattice depth; and the function $R(\eta)$ describes the refractive index profile. In our case $R(\eta)$ describes a couple of one-period low-index defects imprinted in a harmonic lattice and separated by the lattice segment of the length $L_s = nT$ consisting of $n$ lattice periods $T$ [Fig. 1(a)]. In the case of odd separating periods $R(\eta) = 0$ for $nT/2 \leq |\eta| \leq (n+1)T/2$ and $R(\eta) = 1 + \cos(\Omega \eta)$ otherwise; in the case of even number of periods one has $R(\eta) = 0$ for $nT/2 \leq \eta \leq (n/2+1)T$ and $R(\eta) = 1 - \cos(\Omega \eta)$ otherwise, where $\Omega$ is the lattice frequency $\Omega = 2\pi/T$. The parameter $\sigma = \pm 1$ defines the type of nonlinearity (defocusing/focusing). Equation (1) conserves the energy flow $U = \int_{-\infty}^{\infty} |q|^2 \, d\eta$. In view of the scaling properties of Eq. (1), we set $\Omega = 4$ in computer simulations.

The dynamics of linear Bragg-type coupling between low-index defect modes is also far from being trivial, and it exhibits important peculiarities. Thus, we start with a brief summary of the linear analysis. Equation (1) was integrated by a beam propagation method at $\sigma = 0$ with one channel initially excited by a single isolated defect linear mode, which can be found numerically. Such mode features exponentially fading oscillating wings covering a few lattice periods, a property that is typical of linear backward Bragg scattering [20]. One can clearly see (Fig. 1) that in contrast to the behavior exhibited by a conventional coupler, the localization of light in each channel is achieved due to the Bragg-type distributed reflection, and the switching is due to the transmission through the separating lattice segment $L_s = nT$. Figure 1(c) shows typical example of the coupling dynamics. As in a conventional directional coupler, light is slowly tunneled into the second low-index channel. After a certain propagation length, light totally concentrates in the second channel, and then starts backward tunneling to the launching one. Nevertheless, we found that in contrast to conventional couplers, the Bragg-type one operate only in the non-overlapping domains of the lattice depth which are linked to the band-gap structure of the defect-free lattice. These domains are defined by the inequalities $p_{\min}^{(m)} \leq p \leq p_{\max}^{(m)}$ where $m = 1,2,3...$ is the sequential number of finite gap. Notice that in the case of defect-free harmonic lattices, the band-gap structure [Fig. 2 (a)] follows from the properties of a corresponding Mathieu equation. If the lattice depth is outside of the above-mentioned domains, the light beam launched into one of the coupler channels rapidly diffracts [Figs. 1(b) and 1(d)]. To clarify such behavior, we considered the symmetric and anti-symmetric linear modes of the Bragg-type coupler [see Fig. 2(c) for an example]. The propagation constants of these modes $(b_s, b_a)$ belong to the first ($m = 1$) finite gap of the lattice spectrum [Fig. 2(a,b)], and their profiles exhibit exponentially fading oscillating wings [Fig. 2(c)]. The difference of corresponding propagation constants defines the coupling length $L_c = \pi/|b_s - b_a|$ and the light switching might be interpreted as their constructive/destructive interference during propagation.

Figure 2(d) shows the dependencies of propagation constants of symmetric and anti-symmetric modes versus the lattice depth at $L_s = 3T$. When $p$ diminishes ($p \to p_{\min}^{(1)}$) the propagation constant of symmetric mode $b_s$ approaches the *lower* edge of the first finite gap and the mode profile delocalizes transforming itself into a Bloch wave far from the defect. Interestingly, in this point anti-symmetric mode still remains guided. When $p$ grows ($p \to p_{\max}^{(1)}$) and the propagation constant of the anti-symmetric mode approaches the *upper* edge of the first finite gap, this mode in turn delocalizes and transforms into the corresponding Bloch wave. Consequently the coexistence of both modes (which, obviously, is a necessary ingredient for coupling to occur) is possible for the first finite gap only in the finite domain of the lattice modulation depths $p_{\min}^{(1)} \leq p \leq p_{\max}^{(1)}$. For example, at $L_s = 4T$ one has $p_{\min}^{(1)} \approx 0.7$ and $p_{\max}^{(1)} \approx 3.1$ [see the points marked by circles in Fig. 2(d)]. Figure 2(e) shows the monotonic growth of the coupling length with increasing lattice depth. We found that when $n$ is odd the propagation constant of the anti-symmetric mode is higher than that for its symmetric counterpart, while when $n$ is even the situation reverses [Fig. 2(f)]. Clearly, increasing the defects spacing $L_s$ causes an exponential decrease of the coupling strength; also, the coupling length grows rapidly.

A similar scenario holds for modes with propagation constants residing in the second finite gap ($m = 2$). In this case switching is possible for lattice depths $p_{\min}^{(2)} \leq p \leq p_{\max}^{(2)}$, where $p_{\min}^{(2)} \approx 6.2$ and $p_{\max}^{(2)} \approx 11.5$; the linear switching dynamics is shown in Fig. 1(e). Higher-gap switching ($m > 2$) was also found possible in the proper domains of lattice depth. Importantly, Bragg-type coupler can operate using higher-order modes with propagation constants belonging to different gaps. An example of switching with dipole-type mode originating in the second gap is shown in Fig. 1(f), for the lattice depth at which switching with fundamental modes is impossible. It is worth noticing that possibilities of operation in the disjoint domains of lattice depths and switching with higher-order modes are distinctive features of Bragg-type couplers.

Note that while the salient features of the gap solitons have been extensively discussed on the basis of the coupled-mode theory [21], such analysis is based on a decomposition of the beam into forward and backward propagating waves with slowly varying amplitudes, under the condition of Bragg resonance, and the derivation a system of coupled nonlinear equations for those amplitudes. However, our more general approach that takes into account rapid variations of amplitudes, the overlapping of the wave packets in certain spectral domains and also strong coupling, is based on the direct numerical integration of nonlinear Schrödinger equation (1). Here we study the nonlinear response of the Bragg-type coupler by integrating Eq. (1) for different energy flows and input conditions. For simplicity we restrict ourselves to the modes from the first finite gap. As expected, at low energy flows the device acts as a linear Bragg coupler where almost all light is switched between two channels [Figs. 3(a) and 3(c)]. However, with growth of the energy flow $U$, light exhibits incomplete switching and lesser fraction of input energy transmits to the second channel. Figures 3(b) and 3(d) shows the corresponding dynamics for high-energy inputs in focusing and defocusing media, respectively. With further growth of the input energy flow, one arrives at the situation when light remains trapped in the input channel.

Figures 4(a) and 4(c) show the fractional output energy flow (normalized to the input one) trapped in both channels as a function of the input energy flow for focusing and defocusing nonlinearities, respectively. Upon calculation of such switching characteristics, we have fixed the length of the device equal to the linear coupling length $L_c$ and varied the input energy flow. In all cases the input beam profile resembles the profile of single-channel linear guided mode. At low energy flow levels, most of the light is coupled to the output channel and the coupling length is almost the same as linear one. If the energy flow grows the coupling length increases [Figs. 4(b) and 4(d)]. At a threshold input energy flow the coupler enters the regime when energy is redistributed equally between two channels at the output. Thus, for such energy flow the coupling length tends to infinity. Further increasing of the input energy results in defect soliton formation. The excess input energy in this regime goes to the second channel leading to the appearance of weak coupling [Figs. 4(b) and 4(d)].

Interestingly we observed that at $L_s = 4T$, the switching energy in defocusing medium is about 50% bigger than that in focusing one. This remarkable difference between focusing and defocusing media, obtained by direct integration of the governing equation, is in apparent contradiction with the simple model of coupled modes, which predicts equal switching powers for focusing and defocusing cases. To clarify the situation, we simulated the switching with growing channels spacing $L_s$, and encountered that the difference between switching energies tends to zero only when $L_s \to \infty$. For instance at $L_s = 8T$ this difference drops off to $4.3\,\%$. This feature is linked to much slower (in comparison with the standard coupler) fading of the extended oscillating wings of the modes. Thus, for bandgap guidance the coupled-mode theory should be applied with care since it is essentially based on the assumption of weak coupling.

In conclusion, we exposed the main features of nonlinear light switching between two low-index defect channels in periodic lattices forming a Bragg-type coupler. The coupling length of such photonic bandgap device is effectively controlled by the lattice depth and by the channel spacing. In the nonlinear regime the Bragg-type coupler functions as an all-optical switch, and exhibits remarkable difference of switching energies for focusing versus defocusing nonlinearity.

# Figure captions

Figure 1.  (a) Bragg coupler with separation of $L_\mathrm{s} = 4T$ between low-index guiding channels. Linear switching with fundamental modes at (b) $p = 0.6$, (c) $2$, (d) $4$, and (e) $8$. (f) Linear switching with dipole-type mode at $p = 6$. In all cases the right channel of coupler was excited.

Figure 2.  (a) Floquet-Bloch spectrum of uniform lattice. Gray regions show bands and white regions correspond to gaps. (b) Propagation constants of Bloch waves versus transverse momentum $k$ at $p = 1$. Locations of propagation constants of symmetric ($b_\mathrm{s} = -0.974$) and antisymmetric ($b_\mathrm{a} = -0.803$) modes of Bragg coupler with $L_\mathrm{s} = 3T$ are indicated with short black and red lines in (b). (c) Profiles of symmetric (black line) and antisymmetric (red line) linear modes supported by Bragg coupler at $p = 2$ and $L_\mathrm{s} = 8T$. (d) Propagation constants of symmetric and antisymmetric modes and (e) coupling length in low-power limit versus $p$ at $L_\mathrm{s} = 3T$. (f) Propagation constants versus number of lattice periods between cores of coupler at $p = 2$. Lines in (f) are guides for eye.

Figure 3.  Dynamics of switching in Bragg coupler in focusing (a),(b) and defocusing (c),(d) medium. The input energy flows are $U_\mathrm{in} = 0.197$ (a), $0.339$ (b), $0.308$ (c), and $0.481$ (d). In all cases $p = 2$, $L_\mathrm{s} = 4T$.

Figure 4.  (a) Normalized output energy flows concentrated in input and output channels of coupler and (b) coupling length versus input energy flow at $p = 2$, $L_\mathrm{s} = 4T$ for the case of focusing nonlinearity. (c) and (d) show the same, but for the case of defocusing nonlinearity.

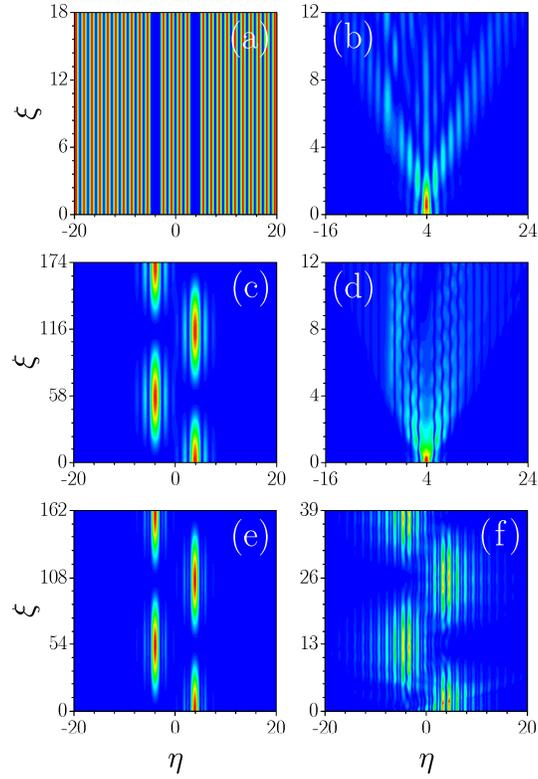

Figure 1. (a) Bragg coupler with separation of $L_s = 4T$ between low-index guiding channels. Linear switching with fundamental modes at (b) $p = 0.6$, (c) $2$, (d) $4$, and (e) $8$. (f) Linear switching with dipole-type mode at $p = 6$. In all cases the right channel of coupler was excited.

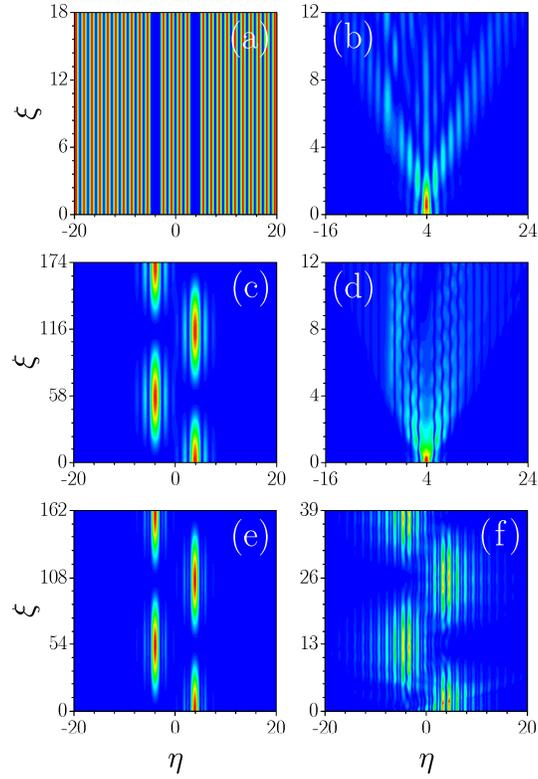

Figure 1. (a) Bragg coupler with separation of $L_s = 4T$ between low-index guiding channels. Linear switching with fundamental modes at (b) $p = 0.6$, (c) $2$, (d) $4$, and (e) $8$. (f) Linear switching with dipole-type mode at $p = 6$. In all cases the right channel of coupler was excited.

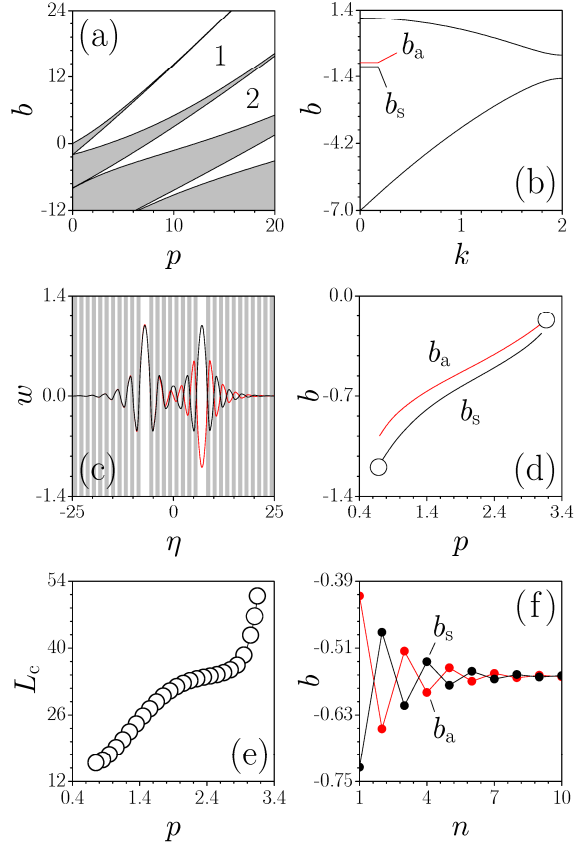

Figure 2. (a) Floquet-Bloch spectrum of uniform lattice. Gray regions show bands and white regions correspond to gaps. (b) Propagation constants of Bloch waves versus transverse momentum $k$ at $p=1$. Locations of propagation constants of symmetric ($b_\mathrm{s}=-0.974$) and antisymmetric ($b_\mathrm{a}=-0.803$) modes of Bragg coupler with $L_\mathrm{s}=3T$ are indicated with short black and red lines in (b). (c) Profiles of symmetric (black line) and antisymmetric (red line) linear modes supported by Bragg coupler at $p=2$ and $L_\mathrm{s}=8T$. (d) Propagation constants of symmetric and antisymmetric modes and (e) coupling length in low-power limit versus $p$ at $L_\mathrm{s}=3T$. (f) Propagation constants versus number of lattice periods between cores of coupler at $p=2$. Lines in (f) are guides for eye.

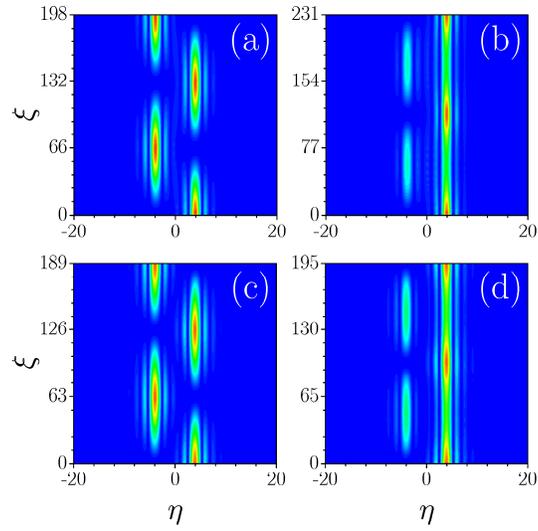

Figure 3. Dynamics of switching in Bragg coupler in focusing (a),(b) and defocusing (c),(d) medium. The input energy flows are $U_{\rm in} = 0.197$ (a), $0.339$ (b), $0.308$ (c), and $0.481$ (d). In all cases $p = 2$, $L_{\rm s} = 4T$.

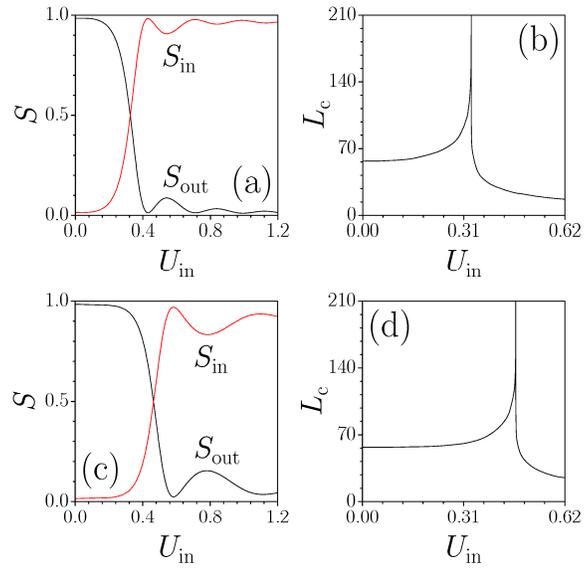

Figure 4. (a) Normalized output energy flows concentrated in input and output channels of coupler and (b) coupling length versus input energy flow at $p = 2$, $L_s = 4T$ for the case of focusing nonlinearity. (c) and (d) show the same, but for the case of defocusing nonlinearity.